\documentclass[runningheads]{llncs}
\usepackage[T1]{fontenc}

\usepackage{graphicx}

\usepackage{euscript}
\usepackage{amssymb}
\usepackage{multirow}
\usepackage[misc]{ifsym}
\usepackage[super]{nth}
\usepackage{diagbox}
\usepackage{hhline}
\usepackage{subcaption}
\usepackage[hidelinks]{hyperref}

\usepackage[table]{xcolor}
\usepackage{ulem}

\newcommand{\boldparagraph}[1]{\paragraph{{\textbf{#1}}}}

\usepackage{xcolor,colortbl}
\usepackage{colortbl}
\usepackage{arydshln}

\usepackage{dcolumn}
\newcolumntype{C}{D{.}{.}{2.2}}

\usepackage{glossaries}

\newacronym{gdloss}{GD loss}{Generalized Dice Loss}
\newacronym{celoss}{CE loss}{Cross-Entropy Loss}

\newacronym[plural=CNNs,firstplural=Convolutional Neural Networks (CNNs)]{cnn}{CNN}{Convolutional Neural Network}
\newacronym{cca}{CCA}{connected component analysis}
\newacronym{mas}{MAS}{Multi-Atlas Segmentation}
\newacronym{mcd}{MCD}{Monte-Carlo Dropout}

\newacronym{iid}{i.i.d.}{independent and identically distributed}
\newacronym{tl}{TL}{Transfer Learning}
\newacronym{da}{DA}{Domain Adaptation}
\newacronym{dg}{DG}{Domain Generalization}
\newacronym{dr}{DR}{Domain Randomization}
\newacronym{sfda}{SFDA}{Source-Free Domain Adaptation}
\newacronym{tta}{TTA}{Test-time Adaptation}

\newacronym{ct}{CT}{Computed Tomography}
\newacronym{mr}{MR}{Magnetic Resonance}
\newacronym{hu}{HU}{Hounsfield Units}

\newacronym{miccai}{MICCAI}{International Conference on Medical Image Computing and Computer-Assisted Intervention}
\newacronym{mmwhs}{MMWHS}{Multi-Modality Whole Heart Segmentation}

\newacronym[plural=ROIs,firstplural=regions of interest (ROI)]{roi}{ROI}{region of interest}
\newacronym[plural=GPUs,firstplural=graphics processing units (GPU)]{gpu}{GPU}{graphics processing unit}

\newacronym{lv}{LV}{left ventricle}
\newacronym{rv}{RV}{right ventricle}
\newacronym{la}{LA}{left atrium}
\newacronym{ra}{RA}{right atrium}
\newacronym{myo}{MYO}{myocardium}
\newacronym{aa}{AA}{ascending aorta}
\newacronym{pa}{PA}{pulmonary artery}

\newacronym{dsc}{DSC}{Dice Similarity Coefficient}
\newacronym{gdsc}{G-DSC}{Generalized Dice Similarity Coefficient}
\newacronym{assd}{ASSD}{Average Symmetric Surface Distance}
\newacronym{hd}{HD}{Hausdorff Distance}

\newacronym{acc}{ACC}{Accuracy}
\newacronym{sen}{SEN}{Sensitivity}
\newacronym{spe}{SPE}{Specificity}

\newacronym{rsc}{RSC}{representation self-challenging}

\newacronym[plural=LA-CaRe-CNNs,firstplural=Left Atrial Cascading Refinement CNNs (LA-CaRe-CNN)]{lacarecnn}{LA-CaRe-CNN}{Left Atrial Cascading Refinement CNN}

\newacronym{lge}{LGE}{late gadolinium enhanced}

\newacronym{myosaiq}{MYOSAIQ}{Myocardial Segmentation with Automated Infarct Quantification}
\newacronym{fimh}{FIMH}{International Conference on Functional Imaging and Modeling of the Heart}

\newacronym{mi}{MI}{myocardial infarction}
\newacronym{af}{AF}{atrial fibrillation}

\newacronym{cdcblock}{CDC block}{convolution-dropout-convolution block}

\newcommand{\lossbase}{L}
\newcommand{\lossdice}{\lossbase_{\text{GD}}}

\newcommand{\imagelettersmall}{x}
\newcommand{\groundtruthlettersmall}{y}

\newcommand{\predictionlettersmall}{p}

\newcommand{\image}{\mathbf{\imagelettersmall}}
\newcommand{\groundtruth}{\mathbf{\groundtruthlettersmall}}

\newcommand{\prediction}{\mathbf{\hat{\groundtruthlettersmall}}}

\newcommand{\predlettersmall}{\mathbf{\hat{\predictionlettersmall}}}
\newcommand{\predstageone}{\predlettersmall_1}
\newcommand{\predstagetwo}{\predlettersmall_2}

\newcommand{\labelpredstageone}{\mathbf{\hat{\groundtruthlettersmall}}_1}
\newcommand{\labelpredstagetwo}{\mathbf{\hat{\groundtruthlettersmall}}_2}

\newcommand{\gtstageone}{\mathbf{\groundtruthlettersmall}_1}
\newcommand{\gtstagetwo}{\mathbf{\groundtruthlettersmall}_2}

\newcommand{\modelweightsbase}{\theta}
\newcommand{\modelweightsstageone}{{\modelweightsbase}_1}
\newcommand{\modelweightsstagetwo}{{\modelweightsbase}_2}

\newcommand{\modelbase}{\mathcal{M}}
\newcommand{\modelstageone}{{\modelbase}_1}
\newcommand{\modelstagetwo}{{\modelbase}_2}

\newcommand{\lossfactorbase}{\lambda}
\newcommand{\lossfactorone}{{\lossfactorbase}_1}
\newcommand{\lossfactortwo}{{\lossfactorbase}_2}

\usepackage{color}

\begin{document}
\title{LA-CaRe-CNN: Cascading Refinement \\ CNN for Left Atrial Scar Segmentation}

\titlerunning{Cascading Refinement CNN for Left Atrial Scar Segmentation}

\author{
Franz Thaler\inst{1,2}\orcidID{0000-0002-6589-6560} \and
Darko \v{S}tern\inst{3}\orcidID{0000-0003-3449-5497} \and
\\
Gernot Plank\inst{1}\orcidID{0000-0002-7380-6908} \and
Martin Urschler\inst{4}\orcidID{0000-0001-5792-3971}
}

\authorrunning{F. Thaler et al.}

\institute{
Gottfried Schatz Research Center: Medical Physics and Biophysics, \\Medical University of Graz, Graz, Austria \and
Institute of Computer Graphics and Vision, Graz University of Technology, \\Graz, Austria \and
AVL List GmbH, Graz, Austria \and
Institute for Medical Informatics, Statistics and Documentation, \\Medical University of Graz, Graz, Austria\\
}

\maketitle             

\begin{abstract}
Atrial fibrillation (AF) represents the most prevalent type of cardiac arrhythmia for which treatment may require patients to undergo ablation therapy.
In this surgery cardiac tissues are locally scarred on purpose to prevent electrical signals from causing arrhythmia.
Patient-specific cardiac digital twin models show great potential for personalized ablation therapy, however, they demand accurate semantic segmentation of healthy and scarred tissue typically obtained from late gadolinium enhanced (LGE) magnetic resonance (MR) scans.
In this work we propose the Left Atrial Cascading Refinement CNN (LA-CaRe-CNN), which aims to accurately segment the left atrium as well as left atrial scar tissue from LGE MR scans.
LA-CaRe-CNN is a 2-stage CNN cascade that is trained end-to-end in 3D, where Stage~1 generates a prediction for the left atrium, which is then refined in Stage~2 in conjunction with the original image information to obtain a prediction for the left atrial scar tissue.
To account for domain shift towards domains unknown during training, we employ strong intensity and spatial augmentation to increase the diversity of the training dataset.
Our proposed method based on a 5-fold ensemble achieves great segmentation results, namely, $89.21$\% DSC and $1.6969$ mm ASSD for the left atrium, as well as $64.59$\% DSC and $91.80$\% G-DSC for the more challenging left atrial scar tissue.
Thus, segmentations obtained through LA-CaRe-CNN show great potential for the generation of patient-specific cardiac digital twin models and downstream tasks like personalized targeted ablation therapy to treat AF.
\end{abstract}

\section{Introduction}

The most prevalent type of cardiac arrhythmia is represented by \gls{af}, which negatively impacts heart function and greatly increases the mortality rate due to an increased risk of stroke~\cite{go2001prevalence,andrade2014clinical}.
Moreover, due to the aging population worldwide, \gls{af} has an increasing incidence rate~\cite{rahman2014global,lippi2021global}.
Radiofrequency ablation is a common therapy for patients with \gls{af}, where cardiac tissues are purposefully scarred to prevent electrical signals from causing irregular heartbeats, aiming to normalize the heart rhythm.
However, one remaining challenge of ablation therapy is the high recurrence rate of more than 40\%, which may require patients to undergo multiple such interventions to successfully overcome the condition~\cite{oral2007radiofrequency}.
A promising direction that aims to increase the success rate of ablation therapy is the generation of cardiac twin models~\cite{corral2020digital,gillette2021framework,li2024towards}, which allow electrophysiological simulation and in turn personalized therapy planning~\cite{boyle2019computationally,Campos2022}.
The generation of accurate cardiac digital twin models of the patient’s anatomy relies on accurate delineations of the anatomical structures of interest, namely, the healthy and scarred tissue.
Such delineations are typically obtained from \gls{lge} \gls{mr} scans~\cite{li2022medical}, where the contrast agent accumulates in scarred tissue, thus allowing its visualization~\cite{selvanayagam2004value}.
However, accurate and efficient analysis of \gls{lge} \gls{mr} images to characterize tissue viability remains challenging due to the thin myocardial walls, complex patterns of scars and limited image quality~\cite{li2020atrial}.

Recent advancements in machine learning contributed to \glspl{cnn} being well-established methods in medical applications like the detection of diseases in medical images~\cite{Esteva2017}, or image segmentation of the vertebrae~\cite{Payer2020-yv}, or the heart~\cite{Payer2018,chen2020deep}.
In literature, several strategies that aim to semantically segment scar tissue from cardiac \gls{lge} scans have been proposed, e.g. to segment the left atrium and atrial scar tissue~\cite{chen2018multiview,yang2020simultaneous,li2022atrialjsqnet}, or to segment the left ventricle and myocardial scar tissue~\cite{chen2022automatic,xu2022bmanet,thaler2024care}.
However, one remaining general challenge of machine learning algorithms that can also be identified when segmenting left atrial scar tissue is domain shift~\cite{li2021atrialgeneral,li2022atrialjsqnet}, which refers to an underperformance in case training and test data are not \gls{iid}.
Domain shift is a consequence of the \gls{iid} assumption of machine learning algorithms, which, if violated, results in higher test errors~\cite{Ben-David2006-ck,Torralba2011-gm} and may even lead to complete model failure~\cite{AlBadawy2018-gi,Pooch2020-vz}.
\gls{dg} refers to a group of approaches that aim to overcome domain shift towards domains that are previously unknown during training, of which one popular group of approaches relies on strong data augmentation~\cite{Zhou2023-jk,Wang2023-vz}.

In this work we propose the \gls{lacarecnn}, a \gls{cnn} cascade inspired by~\cite{thaler2024care} that aims to segment the left atrium as well as left atrial scar tissue from \gls{lge} \gls{mr} scans.
\gls{lacarecnn} is fully 3D, trained end-to-end and consists of a 2-stage \gls{cnn} cascade, where a prediction of the left atrium is obtained from the first stage and later refined to identify areas of scarred tissue.
Aiming to alleviate domain shift towards previously unknown domains, we resample all data to have a consistent and isotropic physical resolution in 3D and employ strong spatial and intensity augmentation to enhance the feature representations observed during training.
Our method is a contribution to the LAScarQS++ track of the CARE2024 Challenge\footnote{CARE2024 Challenge website: \url{http://www.zmic.org.cn/care_2024/}, last accessed in September, 2024} and is externally evaluated by comparing to other contributions to this challenge.

\begin{figure*}[t]
\includegraphics[width=\textwidth]{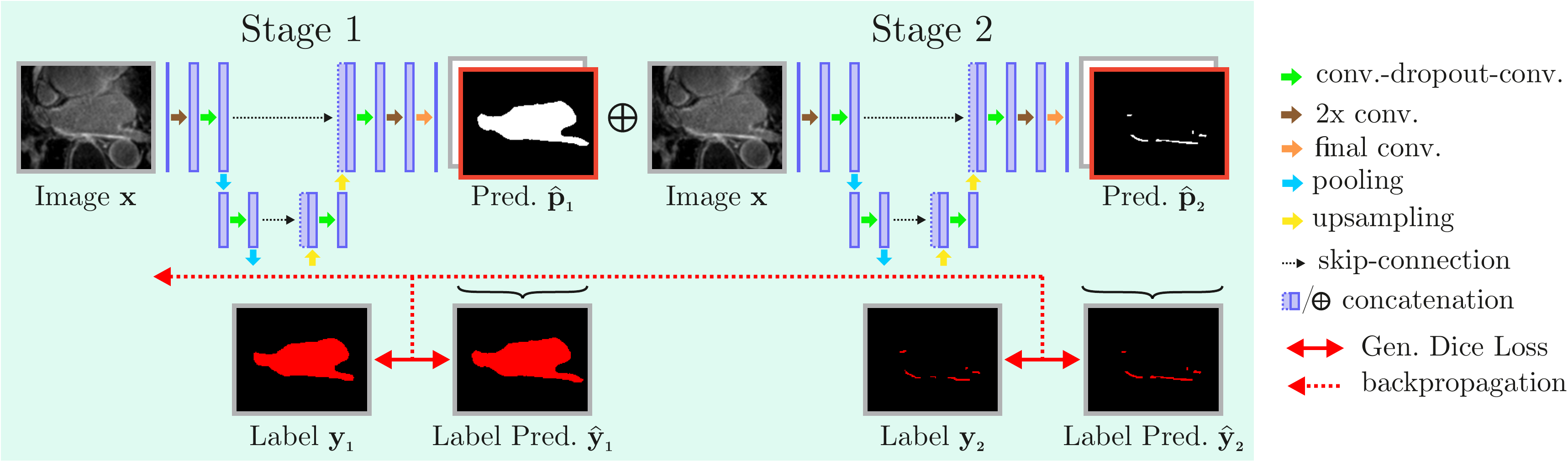}
\caption{
Overview of the proposed LA-CaRe-CNN, a 2-stage CNN cascade to semantically segment the left atrium and left atrial scar tissue from LGE MR scans in 3D.
Stage~1 of LA-CaRe-CNN generates a prediction for the left atrium, which is then concatenated with the original image information in channel dimension before being forwarded to Stage~2.
Predictions for left atrial scar tissue are obtained in Stage~2 by refining the Stage~1 prediction of the left atrium.
}
\label{fig:overview}
\end{figure*}

\section{Method}

In this work we propose \gls{lacarecnn}, a cascading refinement \gls{cnn} that processes \gls{lge} \gls{mr} data in 3D to semantically segment the left atrium and left atrial scar tissue.
\gls{lacarecnn} is closely related to our simultaneous MyoPS++ track submission named MS-CaRe-CNN, which demonstrates the generality of our cascading refinement \gls{cnn} strategy~\cite{thaler2024multi}.
An overview of \gls{lacarecnn} is provided in Fig.~\ref{fig:overview}.

\boldparagraph{Left Atrial Cascading Refinement CNN:}

The network architecture of \gls{lacarecnn} is implemented as a 2-stage \gls{cnn} cascade, which employs two consecutive 3D U-Net-like architectures~\cite{ronneberger2015u} that are trained at the same time in an end-to-end manner. 
For each iteration during training, an image $\image$ with a corresponding ground truth segmentation $\groundtruth$ is randomly sampled from the training set.
At Stage~1 of \gls{lacarecnn}, the image $\image$ is provided as the sole input to the Stage~1 model $\modelstageone(\cdot)$ that generates a prediction $\predstageone$ of the left atrium, without distinguishing healthy and scar tissue.
Formally, this can be expressed as:
\begin{equation}
    \predstageone = \modelstageone (\image ; \modelweightsstageone),
\end{equation}
where $\modelweightsstageone$ refers to the trainable weights of Stage~1 and $\predlettersmall$ is defined as the model output obtained before computing any activation function.
Stage~2 of \gls{lacarecnn} is designed to predict left atrial scar tissue by refining the prediction $\predstageone$ of the left atrium.
However, in order to allow refinement under consideration of the original image intensity information, we concatenate the prediction $\predstageone$ with the image $\image$ in channel dimension before proceeding to Stage~2.
Now, the Stage~2 model $\modelstagetwo(\cdot)$ can be defined as:
\begin{equation}
    \predstagetwo = \modelstagetwo (\predstageone \oplus \image ; \modelweightsstagetwo),
\end{equation}
where $\modelweightsstagetwo$ refers to the trainable weights of Stage~2 and $\oplus$ refers to the concatenation operator in channel dimension.
This yields the Stage~2 prediction $\predstagetwo$ of the left atrial scar tissue.

\gls{lacarecnn} is trained in an end-to-end manner, which allows predictions at any stage to influence all weights that precede the output layer of that stage via standard backpropagation.
In order to compute the loss after obtaining the prediction $\predlettersmall$ of any stage during training, we apply the activation function to acquire the label prediction, i.e. $\prediction = \text{softmax} ( \predlettersmall )$.
Lastly, the training objective for the whole cascade can be defined as:
\begin{equation}
\begin{split}
\lossbase &=
\lossfactorone
\underbrace{\lossdice(\gtstageone, \labelpredstageone; \modelweightsstageone)}_{\text{update $\modelstageone$}}
+ \lossfactortwo
\underbrace{\lossdice(\gtstagetwo, \labelpredstagetwo; \modelweightsstageone, \modelweightsstagetwo)}_{\text{update $\modelstageone$ and $\modelstagetwo$}},
\end{split}
\label{eq:mlc_loss}
\end{equation}
where $\lossdice$ is the generalized Dice loss function, $\groundtruth$ refers to the corresponding ground truth segmentation and $\lossfactorone$ and $\lossfactortwo$ represent multiplicative weighting factors that are both set to 1. 
Please note that the LAScarQS++ track of the CARE2024 Challenge consists of two tasks with different datasets for which separate models needed to be trained.
While the Task~1 dataset includes segmentations of the left atrium as well as the left atrial scar tissue, Task~2 only includes segmentations of the left atrium.
Without scar tissue segmentations, our Task~2 model only consists of Stage~1 of our \gls{lacarecnn}.

\boldparagraph{Addressing Domain Shift:}

In this work, we employ a series of data augmentation techniques that are aimed to account for domain shift towards previously unknown domains, which is a popular strategy to achieve \gls{dg}~\cite{Zhou2023-jk,Wang2023-vz}.
To diversify training data by considering potential differences in local cohorts, we introduce variation to orientation, size and morphology of the anatomy of interest via spatial augmentation techniques, namely, translation, rotation, scaling and elastic deformation.
Furthermore, to account for visual differences like intensity ranges, contrast and signal to noise ratio, which are caused e.g. by the scanner model or the acquisition protocol, we also employ intensity augmentation techniques.
Specifically, for each image during training, we sample a random shift and scale parameter, which are then used to globally modify the intensity values. 
Lastly, we modulate intensities per label before forwarding the image to the \gls{cnn}.

Moreover, we identified that for some cases in the LAScarQS++ dataset the original spacing information that was used to capture the data appears to be missing.
Instead, the spacing information for these cases was set to the default value of $1 \text{ mm}$ in 3D.
This is suboptimal, since \gls{lge} \gls{mr} data is typically acquired in an anisotropic manner, where the out-of-plane spacing in mm is significantly larger compared to the in-plane spacing.
As described in more detail in Section~\ref{sec:implementation_details}, our preprocessing includes isotropic resampling of data in 3D and thus, relies on spacing information being available.
Now, directly processing such data in 3D without correcting the spacing information results in images that appear to be squished in the out-of-plane direction which can negatively impact the predictive performance of a 3D \gls{cnn}.
Effectively, missing spacing information can be viewed as a source of domain shift as a machine learning model might underperform on such data.
Consequently, our preprocessing pipeline includes a procedure, with which we approximate missing spacing information aiming to recover the correct 3D shape of the respective \gls{lge} \gls{mr} scan.
First of all, we consider any scan with a spacing information of $1 \text{ mm}$ in 3D as being incorrectly set. 
From the remaining scans $N$ in the training set, we compute the average physical size $\bar{m}$ of the whole scan, i.e. $\bar{m} = \frac{1}{N} \sum_{n = 1}^{N} v_{n} \cdot s_{n}$, where $v$ refers to the size in voxel and $s$ to the spacing information.
Since the size in voxels is always known for any scan, we can compute the approximated spacing by assuming that the physical size captured by any \gls{lge} \gls{mr} scan for such data is roughly the same.
Thus, we approximate the spacing information for a given scan $i$ by computing $s_{i} = \bar{m} / v_{i}$ before resampling the image and providing it to the \gls{cnn}.
For our final submission on the validation and test set, we revert the approximated spacing information to the original values in order to be consistent with the original data.

\section{Experimental Setup}

\subsection{Dataset}

The dataset used in this work is part of the CARE2024 Challenge and provided for the LAScarQS++ track.
The LAScarQS++ dataset encompasses overall 194 \gls{lge} \gls{mr} scans that have been acquired at three medical centers.
While the ground truth segmentation includes labels for the left atrium and the left atrial scar tissue, the latter is only available for 94 scans that have been obtained at center A.
Consequently, the LAScarQS++ track was separated into two tasks, where Task~1 aims to predict both labels, while Task~2 only evaluates the performance of segmenting the left atrium.
The exact numbers of the training, validation and test set per center for both tracks are provided in Table~\ref{tab:dataset}.

\begin{table*}
\centering
\caption{
The number of subjects per center for Task~1 and Task~2 which are included in the training, validation and test set of the LAScarQS++ track of the CARE2024 Challenge.
}
\begin{tabular}{
p{0.2\textwidth}
| >{\centering}p{0.1\textwidth}
| >{\centering}p{0.1\textwidth}
| >{\centering}p{0.1\textwidth}
| >{\centering\arraybackslash}p{0.1\textwidth}
}

& \multicolumn{1}{c|}{Task 1} & \multicolumn{3}{c}{Task 2} \\
Center & A & A & B & C \\
\hline

Training Set & 60 & 130 & -- & -- \\  
Validation Set & 10 & 10 & -- & 10 \\
Test Set & 24 & 14 & 20 & 10 \\

\end{tabular}

\label{tab:dataset}
\end{table*}

\subsection{Implementation Details}
\label{sec:implementation_details}

All \gls{lge} \gls{mr} scans of the training, validation and test set are resampled to a consistent physical size of $0.8 \times 0.8 \times 0.8$ mm before being processed by the network.
Due to increased GPU memory requirements during training, we computed the center position of the left atrium's ground truth segmentation, around which we extract an image of the size $128 \times 128 \times 160$ voxel.
For images of the validation and test set, we extract a larger image size of $192 \times 192 \times 240$ voxel with the same physical size as used during training that is centered at the center point of the whole scan.
The largest dimension of the image size corresponds to the axis that points from the subject's left to the subject's right.

Data augmentation of training data is performed in 3D for which we apply spatial and intensity augmentation~\cite{Payer2018,payer2019integrating}.
For spatial augmentation, we employ translation ($\pm 20$ voxels), rotation ($\pm 0.35$ radians), isotropic scaling ($[0.8, 1.2]$), anisotropic scaling per dimension ($[0.9, 1.1]$) and elastic deformation (eight grid nodes per dimension and deformation values sampled from $\pm 15$ voxels).
Before applying intensity augmentation, the data is robustly normalized, where the \nth{10} and \nth{90} percentile of intensities are linearly normalized to $-1$ and $1$, respectively.
Then, for intensity augmentation, we randomly sample parameters for intensity shift ($\pm 0.2$) and for intensity scaling ($[0.6, 1.4]$).
The augmentation parameters are sampled uniformly within the respective ranges.
Data from the validation and test set only undergoes robust normalization without any data augmentation.
After obtaining predictions for the validation and test set, we perform a connected component analysis, where we apply a dilation to the prediction of the left atrium in 3D, before merging it with the prediction of the scar tissue.
Then, we remove all components from the prediction obtained from our model that are disconnected from that blob in 3D.

\gls{lacarecnn} is constructed as a series of 3D U-Net-like~\cite{ronneberger2015u} network architectures that follow the same structure at each stage, see Fig.~\ref{fig:overview}.
The network of each stage consists of a contracting and an expanding path with skip-connections and employs five levels of depth.
At each level of either path, we use two convolutions with an intermediate dropout layer~\cite{srivastava2014dropout}, which is followed by a max pooling or a linear upsampling layer, respectively.
Before and after each stage, we additionally employ two respectively three convolution layers.
The kernel size and number of filters of intermediate convolution layers is set to $3 \times 3 \times 3$ and $64$, respectively.
The final convolution layer of each stage employs a $1 \times 1 \times 1$ kernel and $2$ filters.
The convolution kernels are initialized using He initialization~\cite{he2015delving} and we use a dropout rate of $0.1$.
As activation function, we employ leaky ReLU~\cite{maas2013rectifier} with a slope of $0.1$ after intermediate convolution layers.
The final convolution layer of each stage is followed by a softmax activation, however, for Stage~1, the softmax activation is only applied when computing the loss.
Adam~\cite{kingma2014adam} serves as the optimizer with a learning rate of $0.0005$.
Lastly, we employ temporal ensembling~\cite{laine2016temporal} of network weights and train for $100,000$ iterations.
We employed a 5-fold ensemble of independently trained \glspl{lacarecnn} for our final submission, where we average the predictions of the individual models to obtain the prediction of the ensemble.
After loading, the average inference time of the whole ensemble for a single image takes roughly 8~seconds for Task~1 on an NVIDIA GeForce RTX 3090, while training a model lasted roughly 27~hours.
Since Task~2 used a separate dataset which only contained ground truth segmentations of the left ventricle, we employed an ensemble of Stage~1 models for which the average inference time after loading took roughly 6~seconds per subject for the whole ensemble.
Training a model for Task~2 lasted roughly 15~hours.

\begin{table*}[t]
\centering
\caption{
Quantitative results on the validation set of the LAScarQS++ track of the CARE2024 Challenge when segmenting left atrial scar tissue (Task~1) and the left atrium blood cavity (Task~2), respectively. 
Task~1 evaluates the Accuracy (ACC), Specificity (SPE), Sensitivity (SEN), Dice Similarity Coefficient (DSC) and Generalized Dice Similarity Coefficient (G-DSC) in percent, while Task~2 assesses DSC in percent as well as Average Symmetric Surface Distance (ASSD) and Hausdorff Distance (HD) in mm.
All scores were obtained through the submission system.
We compare the performance of a single model to the performance of a 5-fold ensemble and provide results with and without approximating the spacing information (AS) for samples for which this information appeared to be missing.
For our final submission, we used the 5-fold ensemble with AS.
The best score per metric is shown in bold.
}
\resizebox{\columnwidth}{!}{%
\begin{tabular}{ l | c | c | c | c | c | c | c | c | c }

\multirow{2}{*}{\textbf{Method}} & \multirow{2}{*}{\textbf{AS}} & \multicolumn{5}{c|}{\textbf{Task 1: LA scar}} & \multicolumn{3}{c}{\textbf{Task 2: LA cavity}} \\
& & ACC ($\uparrow$) & SPE ($\uparrow$) & SEN ($\uparrow$) & DSC ($\uparrow$) & G-DSC ($\uparrow$) & DSC ($\uparrow$) & ASSD ($\downarrow$) & HD ($\downarrow$) \\
\hline

Single Model & & 76.80 & \textbf{100.00} & 63.17 & 63.32 & 91.38 & 88.42 & 1.9046 & 21.0352 \\
5-fold Ensemble & & 77.08 & \textbf{100.00} & 63.52 & 64.25 & 91.69 & 88.80 & 1.8205 & 20.6002 \\

\hline

Single Model & \checkmark & \textbf{77.78} & \textbf{100.00} & \textbf{64.21} & 64.07 & 91.54 & 89.00 & 1.7530 & 17.7480 \\
5-fold Ensemble & \checkmark & 76.78 & \textbf{100.00} & 62.99 & \textbf{64.59} & \textbf{91.80} & \textbf{89.21} & \textbf{1.6969} & \textbf{17.5315} \\

\end{tabular}
}

\label{tab:quantitative_results_task1}
\end{table*}

\begin{figure*}[t]
\includegraphics[width=\textwidth]{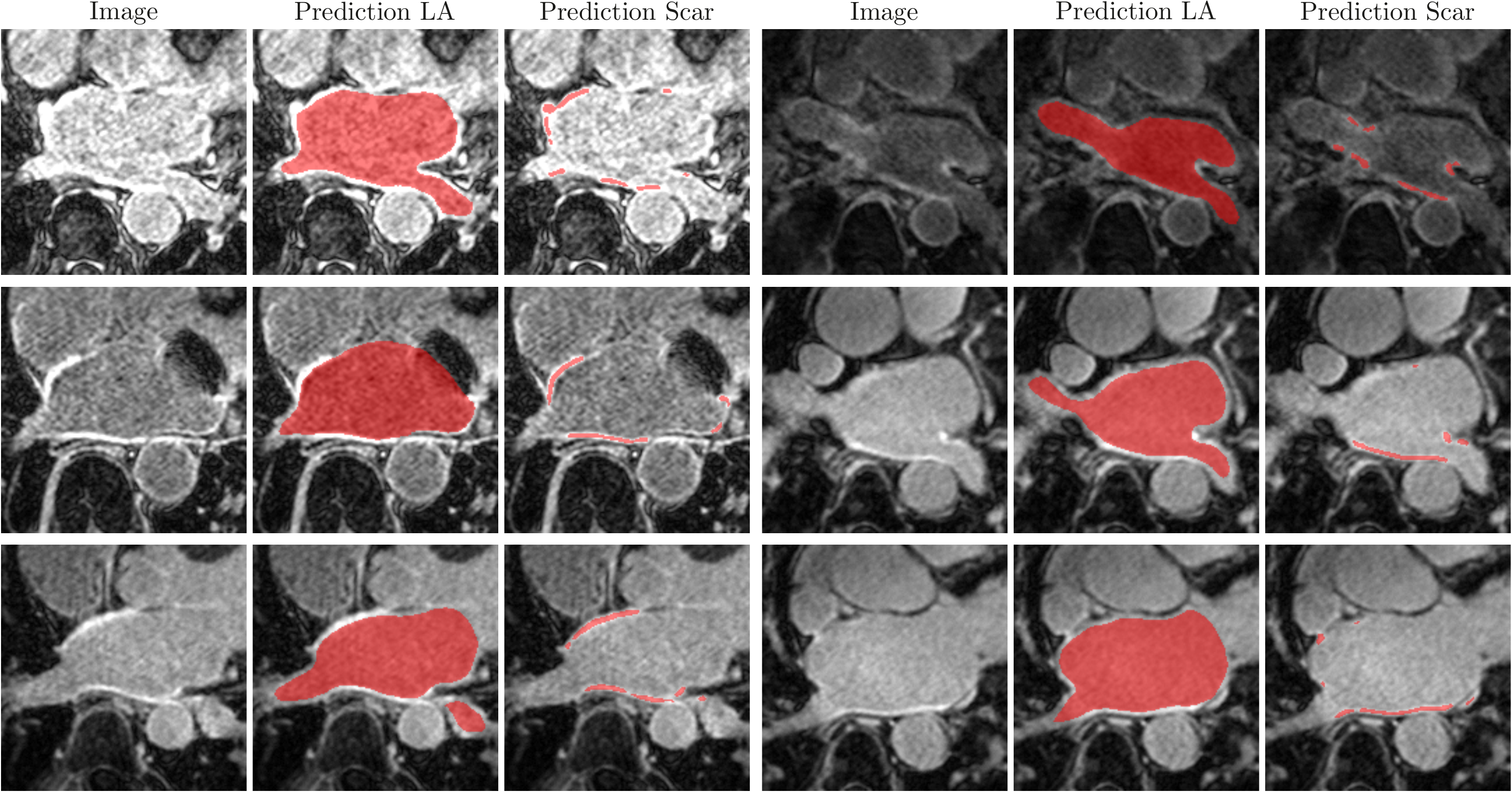}
\caption{
Qualitative results for Task~1 on the validation set of the LAScarQS++ track of the CARE2024 Challenge. 
Shown are exemplary triplets of corresponding images (cols. 1, 4), Stage~1 predictions of the left atrium (cols. 2, 5) as well as Stage~2 predictions of the scar tissue (cols. 3, 6).
}
\label{fig:qualitative_results_task1}
\end{figure*}

\begin{figure*}[t]
\includegraphics[width=\textwidth]{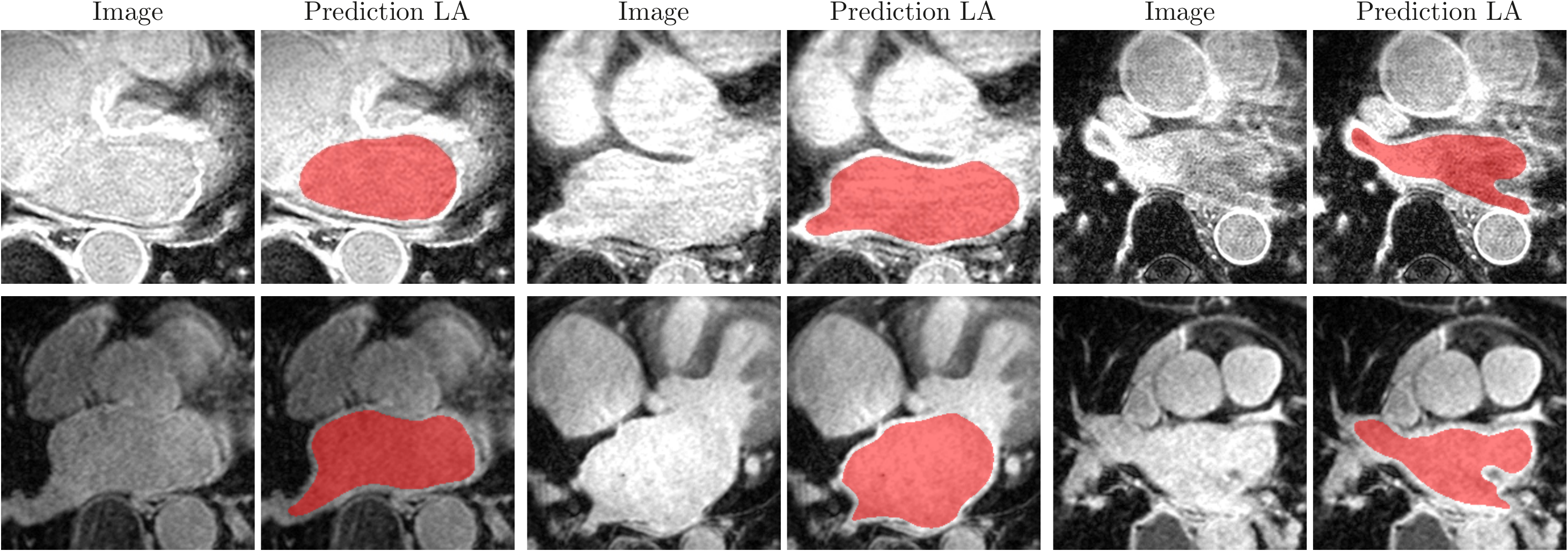}
\caption{
Qualitative results for Task~2 on the validation set of the LAScarQS++ track of the CARE2024 Challenge. 
Shown are exemplary pairs of corresponding images (cols. 1, 3, 5) and predictions of the left atrium (cols. 2, 4, 6).
}
\label{fig:qualitative_results_task2}
\end{figure*}

\section{Results and Discussion}

We evaluate our method on the validation set for which we obtained quantitative scores through the submission system as provided by the CARE2024 Challenge organizers.
In Task~1, quantitative scores are only provided for the prediction of scar tissue which include \gls{acc}, \gls{spe}, \gls{sen}, \gls{dsc} and \gls{gdsc}, all in percent.
The quantitative evaluation of Task~2, which only assesses the prediction of the left atrium, consists of \gls{dsc} in percent, \gls{assd} in mm and \gls{hd} in mm.
The qualitative evaluation is performed by visually inspecting the validation set predictions and comparing them to the original image information, since ground truth labels are not publicly available.
For Task~1, we show the Stage~1 prediction of the left atrium as well as the Stage~2 prediction of the scar tissue.

The quantitative results for both tasks are provided in Table~\ref{tab:quantitative_results_task1}, where we compare the performance of a single model to the performance of a 5-fold ensemble of independently trained models.
Furthermore, we also assess model performance with and without approximating the spacing information (AS) for data of the validation set for which this information appears to be missing.
Specifically, for Task~1, we identified that the spacing information appears to be missing for all images in the validation set, while for Task~2, only 50\% of the validation set were affected.
The quantitative evaluation shows that the score of all metrics for the single model and most metrics for the 5-fold ensemble improved or remained the same when employing AS.
Interestingly, both methods achieved notable improvements on Task~2 even though only 50\% of the validation set were affected by AS.
These results confirm our assumption that approximating the missing spacing information and aiming to recover the 3D shape of the anatomy of interest enables the model to achieve better results by processing the data in a way that is more consistent to the training data.
When comparing the performance of the single model to the 5-fold ensemble both with AS, it can be observed that for Task~1 the single model achieved slightly better scores for \gls{acc} and \gls{sen}, while the 5-fold ensemble resulted in higher \gls{dsc} and \gls{gdsc} scores.
Interestingly, the \gls{spe} score of all compared methods remained the same.
Further, the quantitative evaluation of Task~2 shows that the 5-fold ensemble outperforms the single model in all metrics, which confirms that an ensemble is beneficial due to averaging independent predictions.
Thus, the 5-fold ensemble of independently trained models with AS resembles our final submission.

In the qualitative evaluation on the validation set of Task~1, we show triplets of corresponding images, Stage~1 predictions of the left atrium as well as Stage~2 predictions of the scar tissue, see Fig.~\ref{fig:qualitative_results_task1}.
In general, the semantic segmentations of the left atrium (cols. 2, 5) are very convincing for all cases.
The predictions of the scar tissue also appear to be correct for most cases as they correspond well to regions where the left atrial wall is visually bright due to contrast agents.
However, in some cases, isolated scar predictions consisting of only a small number of pixels can be identified (col 6, row 2 and col 6, row 3), which may require additional assessment by an expert.
Lastly, Fig.~\ref{fig:qualitative_results_task2} provides qualitative results on the validation set of Task~2 for which we show corresponding pairs of images and predictions of the left atrium.
While the validation set of Task~2 consists of data obtained from center~A and center~C, the information which images have been obtained from which center was not disclosed by the organizers.
However, based on the semantic segmentations obtained from our model, we could not identify any cases for which our model did not yield convincing results, which confirms that our model generalizes well to the previously unseen center~C.

\section{Conclusion}

In this work we presented \gls{lacarecnn}, a method that semantically segments \gls{lge} \gls{mr} scans to obtain a prediction for the left atrium as well as left atrial scar tissue.
\gls{lacarecnn} is implemented as a 2-stage \gls{cnn} cascade that processes data in 3D and is trained end-to-end.
By design, our method first predicts the left atrium independently of scar tissue, before refining that prediction in conjunction with the original image information to generate a prediction for the left atrial scar tissue.
Moreover, we aim to address domain shift towards domains unknown during training, by employing strong data augmentation techniques based on intensity and spatial augmentation inspired by \gls{dg} approaches.
Our proposed method consisting of a 5-fold ensemble achieves a score of $89.21$\% \gls{dsc} and $1.6969$ mm \gls{assd} when segmenting the left atrium, as well as $76.78$\% \gls{acc}, $64.59$\% \gls{dsc} and $91.80$\% \gls{gdsc} when segmenting the challenging left atrial scar tissue.
These results confirm that semantic segmentations obtained from our method have great potential for further use in generating patient-specific cardiac digital twin models.
In turn, these digital twin models can be used e.g. for electrophysiological simulation which allows personalized targeted ablation therapy to treat \gls{af}.

\begin{credits}
\subsubsection{\ackname}
This research was funded in whole or in part by the Austrian Science Fund (FWF) 10.55776/PAT1748423 and also by the CardioTwin grant I6540 from the Austrian Science Fund (FWF).

\subsubsection{\discintname}
The authors have no competing interests to declare that are relevant to the content of this article.

\end{credits}

\bibliographystyle{splncs04}
\bibliography{references}

\end{document}